\documentclass[
twocolumn,
 amsmath,amssymb,
 aps, 
]{revtex4-2}

\usepackage{graphicx}% Include figure files
\usepackage{dcolumn}% Align table columns on decimal point
\usepackage{bm}% bold math
\usepackage{placeins}

\usepackage{booktabs}

\begin{document}

%\preprint{APS/123-QED}

\title{\textbf{Non-Linearities In Atomic Radio Receivers: Harmonic And Intermodulation Distortion}
}% 

\author{Lu{\'\i}s~Felipe~Gon\c{c}alves}
\email{Luis@RydbergTech.com}
%\homepage{www.RydbergTechnologies.com}
\affiliation{Rydberg Technologies Inc. Ann Arbor, Michigan 48103 USA}
\thanks{This work was supported by Rydberg Technologies Inc. and, in part, by the Defense Advanced Research Projects Agency (DARPA) SAVaNT Program under agreement No. HR00112190065. Distribution Statement "A" (Approved for Public Release, Distribution Unlimited).  The views and conclusions contained in this document are those of the authors and should not be interpreted as representing the official policies, either expressed or implied, of the U.S. Government.}
 
\author{Teng Zhang}%
%\email{TZhang@RydbergTech.com}
\affiliation{Rydberg Technologies Inc. Ann Arbor, Michigan 48103 USA}%

\author{Georg Raithel}
%\email{Raithel@Rydbergtech.com}
\affiliation{Rydberg Technologies Inc. Ann Arbor, Michigan 48103 USA}%

\author{David A. Anderson}
\email{Dave@RydbergTech.com}
\affiliation{Rydberg Technologies Inc. Ann Arbor, Michigan 48103 USA}%

\date{\today}% It is always \today, today,

\begin{abstract}
Rydberg sensors offer a unique approach to radio frequency (RF) detection, leveraging the high sensitivity and quantum properties of highly-excited atomic states to achieve performance levels beyond classical technologies. Non-linear responses and distortion behavior in Rydberg atom receivers are critical to evaluating and establishing performance metrics and capabilities such as spur-free dynamic range and tolerance to unwanted interfering signals. We report here on the measurement and characterization of non-linear behavior and spurious response of a Rydberg atomic heterodyne receiver. Single-tone and two-tone testing procedures are developed and implemented for measurement of harmonic and inter-modulation distortion in Rydberg atomic receivers based on multi-photon Rydberg spectroscopy and radio-frequency heterodyne signal detection and demodulation in an atomic vapor.  For a predetermined set of atomic receiver parameters and RF carrier wave in the SHF band near-resonant to a cesium Rydberg transition, we measure and characterize atomic receiver selectivity, bandwidth, roll-off,  compression point (P1dB), second-order (IP2) and third-order (IP3) intercepts, and spur-free dynamic range.  Receiver intermodulation distortion is characterized for the case of an interfering signal wave applied at two frequency offsets relative to the near-resonant reference local oscillator, $\Delta F/F= 10^{-4}$ at 6~dB and $10^{-6}$ at 22~dB single-tone bandwidths, respectively.  We observe that under suitable operating conditions the atomic receiver can exhibit a 
suppression of harmonic and inter-modulation distortion relative to that of classical receiver mixer amplifiers.  Finally, we describe how the non-linear behaviors of atomic receivers can provide unique, controllable RF signatures inaccessible by classical counterparts and propose their use to realize secure communication modalities and applications. 

%\begin{description}
%\end{description}
\end{abstract}

\maketitle

%\tableofcontents

\section{Introduction}
Rydberg atom-based radio-frequency (RF) sensors are a rapidly advancing quantum technology, offering transformative capabilities in RF applications ~\cite{adams2019review,Anderson2.2017, Anderson.2019, guo2023eit,AndersonRFMS,kumar2017eit,adams2022enhanced}. These sensors feature ultra-wide frequency coverage spanning from DC to THz ~\cite{Holloway.2014, Miller.2016,adams2024highl} with sub-wavelength sensing elements ~\cite{AndersonGSMM.2018, Holloway2.2014} and capabilities ranging from high-intensity RF field measurement~\cite{Anderson.2014, Paradis.2019} and detection of incoherent RF electromagnetic radiation~\cite{SimonsNoise.2018}, to absolute high-precision field measurements of over-the-air (OTA) RF electric fields at the 0.1~\% level~\cite{Miller.2016} and in situ field measurements of charged plasmas ~\cite{AndersonPlasma.2017}. 
A notable advancement in Rydberg atom-based RF technologies was the development of phase-sensitive Rydberg detectors and receivers that incorporate a fiduciary reference wave into the atomic detection medium, enabling direct Rydberg atom demodulation of amplitude, frequency, and phase of OTA electromagnetic waves ~\cite{Anderson.2019}. Implementations include all-optical RF phase-sensitive detection, where optical RF reference local oscillators interfere with OTA signals in the Rydberg atom state-space, enabling ultra-high RF phase resolution at optical diffraction limits ~\cite{AndersonPhase.2022}, and RF heterodyne reception, utilizing free-space RF reference waves to interfere with OTA signal fields, demonstrating significant enhancement in sensitivity and selectivity in Rydberg heterodyne receivers ~\cite{SimonsPhase.2019, Jing.2020,legaie2023millimeterwave}. Other recent demonstrations of continuous RF tuning using off-resonant and on-resonant couplings~\cite{Anderson.2017,Prajapati.2023} and simultaneous wide multi-band demodulation~\cite{MeyerMulti.2023} exemplify the potential of Rydberg sensors in applications requiring broadband frequency-agility that offer new possibilities in resilient communications and wireless networks.

Leveraging early advances, atomic RF demodulation and reception of modulated RF signals has lead to the advent of phase-sensitive atomic receivers for analog and digital radio communications ~\cite{Anderson.2019,AndersonAtomicReceiver.2020B, AndersonIEEEAES.2020, Meyerpub.2018, Debpub.2018, HollowayRydbergRadio.2019, Liu.2022, MeyerMulti.2023}. The first Rydberg sensor device, the Rydberg field probe (RFP) and measurement system (RFMS)~\cite{AndersonRFMS} demonstrated atomic RF field measurement, signal waveform reception and non-invasive high-resolution near-field antenna characterizations ~\cite{Cardman.2020}. More recently, the  authors have developed a compact, sub-100-liter volume fieldable atomic receiver (ARx) (highlighted in ~\cite{Schlossberger2024}) demonstrating groundbreaking advances in deployable Rydberg device hardware, and heralding the first demonstration of long-range quantum radio communication using a Rydberg atomic receiver ~\cite{AndersonIEEEURSIARx.2024}. These milestones highlight the potential of Rydberg technology to revolutionize RF communications by leveraging quantum sensing principles for long-distance highly sensitive RF detection.

Despite the remarkable progress made in Rydberg atom-based RF sensors, a critical aspect remains unexplored: the non-linear behavior of Rydberg atom receivers and their harmonic and intermodulation distorition performance in RF signal reception. Non-linear responses and distortion behavior are essential factors in evaluating the capabilities of radio frequency receivers, particularly for determining key metrics such as spur-free dynamic range and resilience to unwanted interfering signals. To date, methods to systematically establish and improve these performance metrics in Rydberg receivers have not been developed. In this work, we address this gap by reporting on measurements and characterizations of non-linear behavior and spurious response in Rydberg atomic RF receivers, shedding light on the fundamental performance advantages and potential optimizations of these quantum sensors.

Two-tone and spurious response testing is commonly performed on classical receiver systems to evaluate their spur-free dynamic range and tolerance to unwanted interfering signals. Receiver performance considerations include protection against radio-frequency (RF)-induced damage to the receiver electronics, the degree of degradation allowed in receiver performance in the presence of strong interfering signals\cite{akaiwa2013interband}, and overall system performance in congested electromagnetic environments.  Here we translate these traditional concepts and testing methods from RF engineering to quantum Rydberg atomic sensors and receivers, and develop analogous testing procedures suitable for atomic receivers. We investigate non-linear effects in the quantum optical Rydberg atom response to RF fields in the receiver front end. Since the physics principles underlying Rydberg atomic RF receivers are fundamentally different from those underlying traditional RF receivers, the presented study of nonlinear behavior in Rydberg atomic RF receivers enables direct performance comparisons of atomic quantum receivers to analogous classical technologies. 

This paper is organized as follows: In Section~\ref{Section1TestSetup}, we describe the RF test setup and Rydberg atomic heterodyne receiver used to conduct the atomic RF measurements and characterization at 10~GHz. In Section~\ref{Section2Selectivity}, we present single-tone testing of the atomic receiver and measurements of receiver selectivity including intermediate-frequency (IF) bandwidth and signal rejection ratio. Section~\ref{Section3HarmonicDist} presents single-tone measurement and characterization of harmonic distortion. In Section~\ref{Section4TwoToneIMD}, we present two-tone tests and measurements of intermodulation distortion in the atomic receiver response for in-band and out-of-band interference signals.  A figure of merit for intermodulation distortion (IMD) performance comparison between quantum and classical receivers is established  to quantify the receiver’s non-linear response to incident electric fields of the RF signal waves in the saturation regime, and used to benchmark performance against a classical low-noise amplifier. In Section~\ref{Section5Discussion}, we provide a discussion of the results, and in Section~\ref{Section6Conclusion} a concluding summary and outlook.

\section{Atomic receiver testing}\label{Section1TestSetup}

\begin{figure}[ht!]
\centering
\includegraphics[width=1\linewidth]{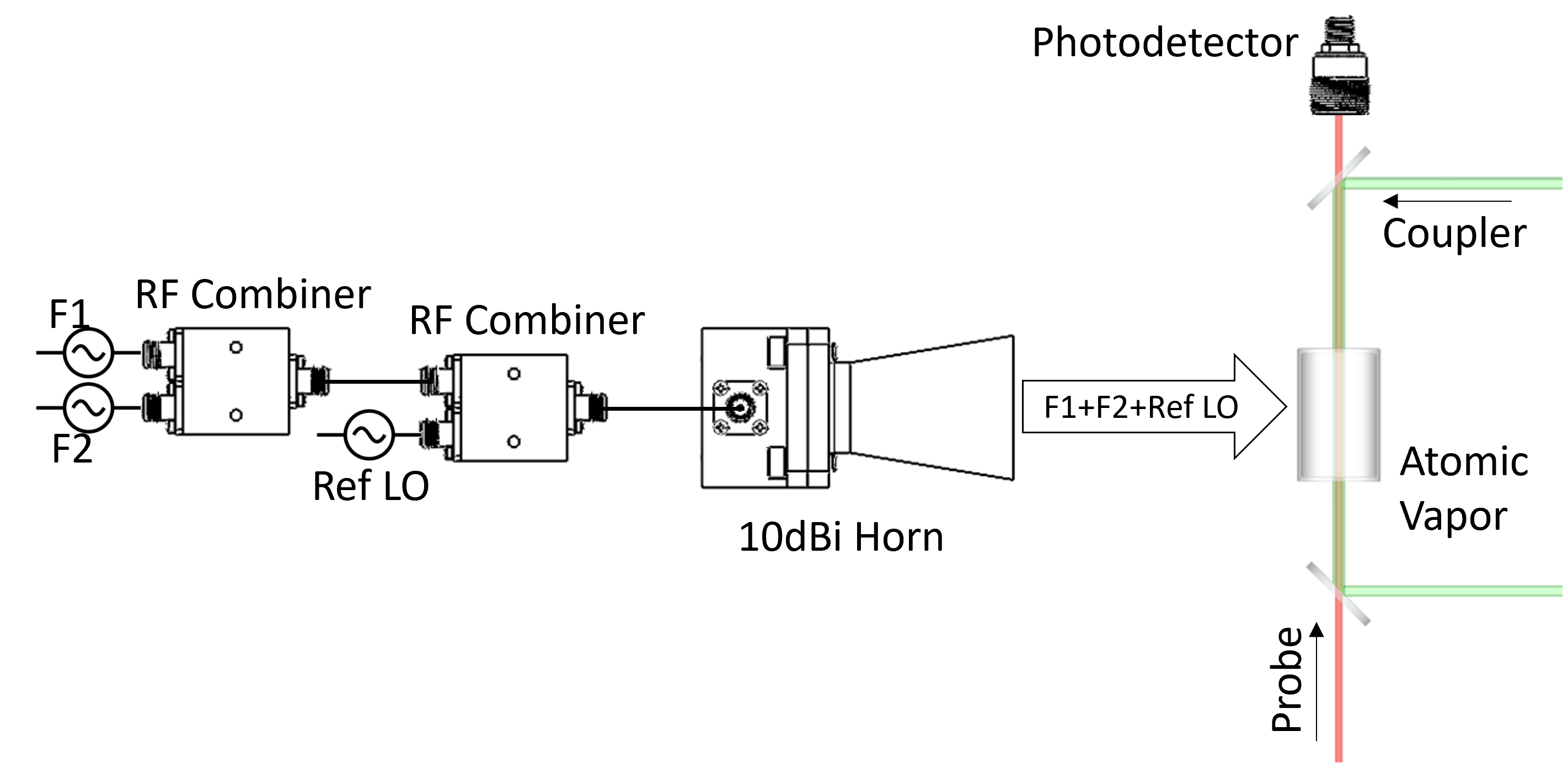}
\caption{Testing setup for a Rydberg atomic receiver.  The $F_1$ or/and $F_2$ tones are combined with the Ref LO and injected into a 10~dBi antenna which transmits the combined RF signals OTA to an atomic vapor.  The response of the RF-sensitive Rydberg atoms in the Cs vapor is measured all optically using a photodetector. The amplified photodetector output is processed using a signal analyzer. For single-tone testing we only apply $F_1$, while for two-tone testing we apply both $F_1$ and $F_2$.}
\label{fig:2tone}
\end{figure}

The test and measurement setup for single-tone and two-tone testing of a Rydberg atomic RF heterodyne receiver is illustrated in Figure~\ref{fig:2tone}.  Two signal tones at frequencies $F_1$ and $F_2$ are generated and combined with the atomic receiver reference local oscillator (LO) at frequency $F_{LO}$ used in phase-sensitive atomic RF heterodyne reception\cite{Jing.2020}. The three frequencies are generated by three different signal generators synchronized to a rubidium clock, allowing independent frequency and power control of each signal tone and the LO.  The combined $F_{1}$+$F_{2}$+$F_{LO}$ waveform is coupled into a WR-75 horn antenna with gain G = 10~dBi, which transmits the waveform OTA to the atomic vapor of the atomic receiver located approximately 18~cm away in the antenna far-field. The atomic vapor is a cesium gas held at room temperature in a glass cell. Multi-photon Rydberg spectroscopy and electromagnetically-induced transparency (EIT) are used to access RF-field sensitive Rydberg states of the cesium gas with two counter-propagating laser beams at 852~nm (probe beam) and 510~nm (coupler beam).

During RF signal reception, both probe and coupler lasers in the atomic receiver are frequency stabilized to linewidths below 1~kHz at target operating points near the Rydberg resonance ~\cite{AndersonAtomicReceiver.2020B} using a combination of atomic and ultra-stable optical references~\cite{legaie2023millimeterwave}. The probe laser frequency is stabilized in the vicinity of a cesium D2 hyperfine transition~\cite{Steck.2023}, and the coupler laser frequency is set to couple the $6P_{3/2} \rightarrow 42D_{5/2}$ transition to access the $42D_{5/2}\rightarrow 43P_{3/2}$ Rydberg RF resonance at $9.9376$~GHz.  The probe beam transmission through the atomic vapor is measured on an amplified photo-detector for absorption measurements of RF-induced changes to the Rydberg atom vapor.  The electronic readout is measured using a spectrum analyzer (SA) for measurement of the atomic receiver response to applied RF signal tones. We perform atomic receiver testing in the SHF-band, and configure the atomic heterodyne receiver LO field frequency to $F_{LO}=9.9376$~GHz, resonant with the $42D_{5/2}\rightarrow 43P_{3/2}$ Rydberg transition.  In the presented measurements, either the $F_1$ or both $F_1$ and $F_2$ signal tones are applied to the atomic receiver for single-tone and two-tone measurements, respectively. The atomic heterodyne receiver response to the applied tones is measured on the SA at intermediate frequencies $f_1=F_{LO}-F_1$ and $f_2=F_{LO}-F_2$, as well as at overtones and mixing products of $f_1$ and $f_2$. All RF fields applied to the atomic receiver are calibrated to absolute electric field levels by performing atom-based RF electric field measurements using RF-induced Autler-Townes splittings or AC Stark shifts of Rydberg levels of the receiver atoms. The calibration measurements are performed at test fields that are sufficiently high to generate high-fidelity Autler-Townes splittings.

\section{Single-tone testing: Selectivity, IF Bandwidth, and filter roll-off}\label{Section2Selectivity}

We first perform a single-tone measurement to characterize the atomic receiver  selectivity, IF bandwidth, and filter roll-off for an SHF-band signal tone following the procedure described in our earlier work~\cite{legaie2023millimeterwave}. The response of the atomic receiver to a tone at frequency $F$ and field amplitude $E_F=0.017$~V/m($-35.4$~dBV/m) is measured as a function of its frequency detuning in the IF $f= F-F_{LO}$ relative to the receiver LO frequency $F_{LO}=9.9376$~GHz resonant with the Rydberg transition with amplitude $E_{LO}=0.21$~V/m($-13.6$~dBV/m). Throughout this paper, we use dBV/m as well as V/m as the electric field units. The log scale unit dBV/m is better to visualize the scaling in the plots.

The atomic receiver differs from conventional receivers in that the incident signal is an over-the-air (OTA) RF electric field that is impingement onto the atoms. The atomic receiver may or may not have an antenna-like structure attached to it; in any case the atoms sense an RF electric field, as opposed to an RF voltage. The RF electric field is transmuted by the atoms into an optical signal, which is processed downstream via photodetectors and amplifiers. In contrast, a traditional system has an antenna that converts the incident OTA RF electric field into an RF voltage, which constitutes the input of the receiver. The receiver is considered separate from the antenna. An important distinction between an atomic and a conventional receiver therefore is that the atomic receiver’s input is an electric field, while the conventional receiver’s input is a voltage, or a power delivered into the receiver’s input impedance. To compare the sensitivities of the two types of receivers, one requires the antenna factor (AF) that relates the voltage at the antenna base to the incident OTA RF electric field. The AF’s dependence on wavelength and antenna gain is well known. While in the present work we are not concerned with sensitivity comparisons, it is important to understand the differences between atomic and conventional receivers to understand the units utilized in the following.

\begin{figure}[ht!]
\centering
\includegraphics[width=0.72\linewidth]{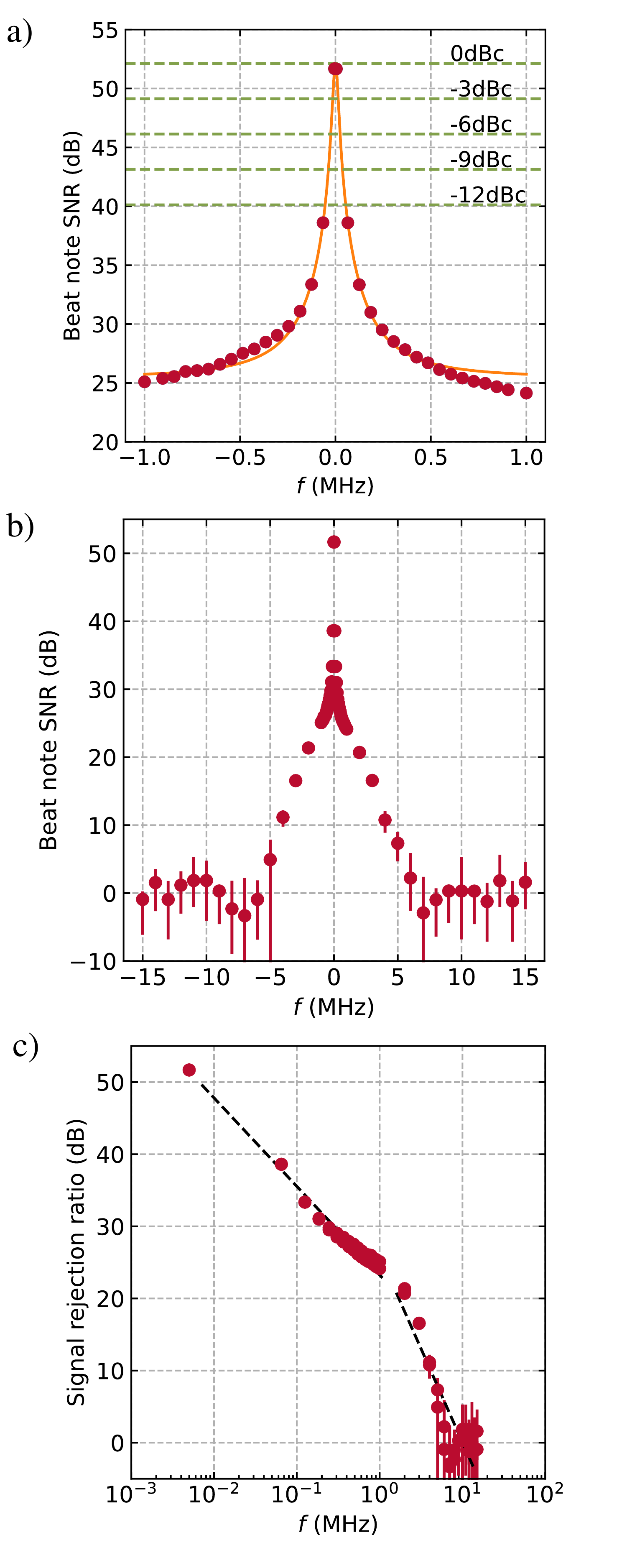}
\caption{Atomic receiver response as function of signal frequency detuning $f=F-F_{LO}$ relative to $F_{LO}=$9.9376~GHz  resonant with the  Cs $42D_{5/2} \rightarrow 43P_{3/2}$ Rydberg transition: a) Atomic receiver IF response signal to noise (SNR) in dB versus $f$ over a range $\pm 1~MHz$.  Empirical fit of the response curve to a Voigt (solid line) gives Gaussian and Lorentzian width of $\Gamma_{G}/2 \pi=$$22.8\pm0.9$~kHz (FWHM = 56.6~kHz) and $\Gamma_{L}/(2 \pi) = 12.6\pm1.6$~kHz, respectively.  The IF bandwidths at 3~dB, 6~dB, 9~dB, and 12~dB, listed in Table~I, are given by the separations between the intersects between the Voigt fit and respective horizontal dashed lines. Error bars are the standard deviation from an average of 3 independent measurements of the beat note signal at each $f$. b) Receiver IF response signal to noise (SNR) in dB versus $f$ over a range $\pm 15$~MHz.  c) Atomic receiver filter roll-off plot of the signal rejection ratio in dBc versus $f$.  Linear fits in two distinct roll-off regimes (dashed lines) yield 10~dB/decade for $f<$1.1~MHz and 30~dB/decade for $f>$1.1~MHz.}
\label{fig:selectivity}
\end{figure}

Henceforth we adopt the notion that the atomic receiver’s input is the incident OTA RF electric field present at the location of the atoms. Throughout the paper, the RF electric field strength, $E$, is measured in units of dBV/m. To convert into SI units, use $E_{SI} = 1~$V/m$ \times 10^{E/20}$, where $E$ is the field in dBV/m. These definitions are used for the $LO$ and for the signal tones $F$, $F_1$ and $F_2$. The receiver’s output power, $P_{out}$, is given by the spectrum-analyzer reading from the amplified photo-detector signal at the IF frequency or frequencies. The output power $P_{out}$ is measured in units of dBm and depends on the transimpedance amplifier gain and other operational parameters. $P_{out}$ is subject to a noise floor. The signal-to-noise ratio (SNR) at the IF is given by the difference between $P_{out}$ at the IF frequency in dBm and the noise floor in dBm. The dBm-level difference between the IF signal and features at other frequencies is measured in units of dBc (suppression in dB relative to the carrier). In our figures, we show incident RF fields in units of dBV/m, while $P_{out}$, SNR and signal rejection ratio (SRR)  are given in units of dBm,  dB, and dBc. In this manner, the scaling laws for non-linear effects as a function of incident RF field can be visualized well. In the text, we sometimes also provide RF field strengths in units V/m to convey the physical electric fields the atoms are exposed to.

Figure~\ref{fig:selectivity}a shows the IF signal response of the receiver to the applied tone F as a function of $f$ for $F$ scanned over approximately $\pm 1$~MHz centered near $F_{LO}$.  To quantify the selectivity, the measured response curve is fit empirically to a Voigt function, from which we obtain the peak signal response at $f=0$ and IF bandwidths at 3, 6, 9, and 12~dB (where the response drops by 3, 6, 9, and 12~dB in power). Measured IF bandwidths are given in Table~\ref{table:if-bandwidth}.

In Figure~\ref{fig:selectivity}c, we show the measured signal rejection ratio (SRR) in dBc relative to the peak signal response for $f$=10~kHz out to $f=\pm 15$~MHz. The dashed black lines are separate linear fits to the data in two regions, in which the SRR exhibits qualitatively different roll-off behavior. In the domain $f=$ 10~kHz to 1~MHz, the signal roll-off approximately follows 10~dB/decade while from 1~MHz to 15~MHz it approximately follows 30~dB/decade.

\begin{table}[h!]
\begin{center}
\caption{IF bandwidth of the atomic receiver at a carrier frequency of 10~GHz}
\label{table:if-bandwidth}
\begin{tabular}{@{}ccccc@{}}
\toprule
 \textbf{\begin{tabular}[c]{@{}c@{}}Carrier Frequency\\ (GHz)\end{tabular}} &
 \textbf{\begin{tabular}[c]{@{}c@{}}3~dB\\ (kHz)\end{tabular}} &
 \textbf{\begin{tabular}[c]{@{}c@{}}6~dB\\ (kHz)\end{tabular}} &
 \textbf{\begin{tabular}[c]{@{}c@{}}9~dB\\ (kHz)\end{tabular}} &
 \textbf{\begin{tabular}[c]{@{}c@{}}12~dB\\ (kHz)\end{tabular}} \\ \midrule
 10     & 61      & 98      & 145     & 212 \\ \bottomrule
%100*    & 127     & 195     & 164     & 343 \\ 
\end{tabular}
\end{center}
\end{table}

\section{Single-tone testing: Harmonic distortion}\label{Section3HarmonicDist}

\begin{figure}[h!]
\centering
\includegraphics[width=0.9\linewidth]{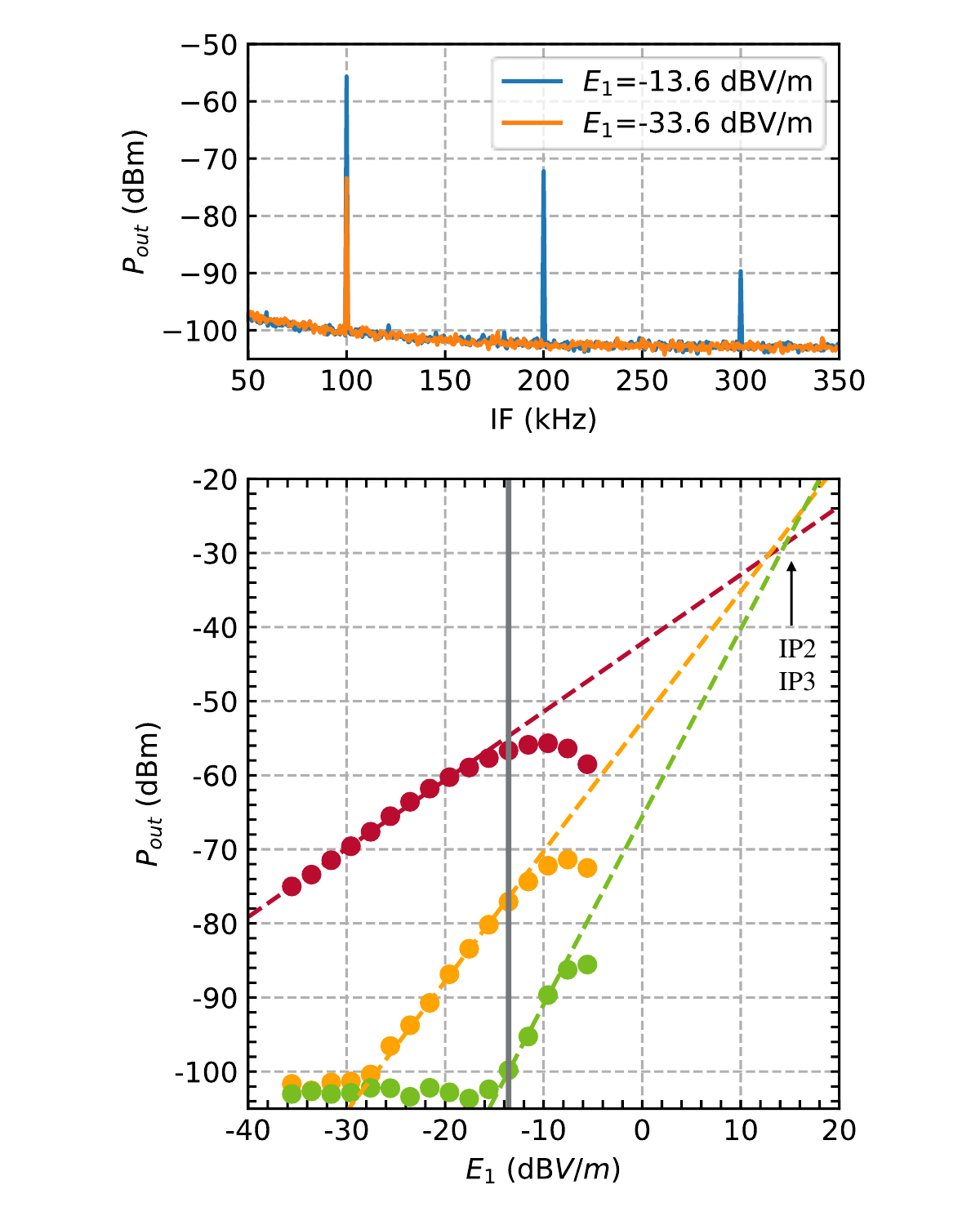}
\caption{Top: Power spectra of the receiver response measured on the SA  for a single tone $F_1$ applied at the vapor cell with electric-field amplitudes $E_{1} = $0.021~V/m (-33.5~dBV/m)(orange) and 0.209~V/m (-13.5~dBV/m)(blue).  Bottom: Measured receiver power for $f_i$ fundamental and higher harmonics i = 1, 2, and 3 as a function of applied tone $E_{1}$ field in units of dBV/m. Linear slopes (dashed lines), the measured $f_1$ P1dB compression point at $E_1= -13.6$~dBV/m (vertical solid line), and $f_2$ and $f_3$ IP2=12.5 and IP3=14.0~dBV/m intercept points are indicated.}
\label{fig:harmonics}
\end{figure}

To investigate the non-linear behavior of the receiver to an incident RF wave, we perform a single-tone measurement of non-linear distortions within the pass-band of the receiver characterized in Figure~\ref{fig:selectivity}. This is performed using the testing setup shown in Figure \ref{fig:2tone}, with an applied tone $F_{1}=$9.9377~GHz that is offset from the F$_{LO}=$9.9376~GHz by $f_1=$+100~kHz approximately at the 6~dB IF bandwidth.  No second tone F$_2$ is applied for this measurement. The measurement is performed at a fixed receiver LO field strength $E_{LO}=$0.21~V/m($-13.6$~dBV/m), and by varying the electric field strength $E_{1}$ of F$_1$ and recording the $f_1$ signal response on the SA.  The top plot in Figure~\ref{fig:harmonics} shows power spectra of the response measured on the SA over a span of 50~kHz to 350 ~kHz for two $F_1$ field strengths. $P_{out}$ is the power measured at the photodetector in the IF domain. At a field strength $E_1=0.02$~V/m($-34.0$~dBV/m), approximately one order of magnitude lower than the receiver's LO field strength, the fundamental $f_1$ beat note is measured at a signal level of 25~dB. Increasing the field strength by 20~dB to $E_1=0.2$~V/m($-14.0$~dBV/m), a level comparable to the receiver's $E_{LO}$ field level, second and third harmonics emerge at frequencies (and signal levels) $f_2=200$~kHz (29.5~dB) and $f_3=300$~kHz (12.8~dB), respectively, as a result of harmonic mixing of the tone and local oscillator fields by the non-linear atomic medium. The bottom plot of  Figure~\ref{fig:harmonics} shows the  received signal power at $f_1$ and the  harmonic overtones $f_2$ and $f_3$ generated by the atomic receiver as a function of electric field $E_1$ of the applied tone $F_1$ in units of dBV/m.  The field is calibrated using the receiver Rydberg atoms.  A linear fit to the measured $f_1$ response over the applied field range $E_1<$-20~dBV/m gives a slope of  $P_{out}$(dBm)/$E_1$(dBV/m)=0.924$\pm$0.016, as expected for the first-order fundamental response, and allowing the designation of the  P1dB compression point at an $F_1$ field strength equal to $E_1$=-13.6~dBV/m. Similarly, we measure linear slopes for the $f_2$ second harmonic and $f_3$ third harmonic overtones to be 1.75$\pm$0.05 and 2.8$\pm$0.15, respectively, in good agreement with expected values of 2 and 3.  The second order intercept (IP2) and third order intercept (IP3) of the harmonic distortion products are also obtained, and nearly coincide at the $F_1$ field strength of $E_1$=12.5~dBV/m.

While the results in Fig.~3 mimic those of conventional receivers, it must be stressed that linearity and nonlinear distortion of the atomic receiver are not due to the characteristics of electronic components but are due to the high-order electric-field mixing behavior of the atomic medium. The resemblance of the scaling found for the atomic receiver with the scaling expected for conventional receivers, including the near-linear response at the fundamental IF frequency, presents a valuable finding. We note that, at P1dB in Fig. 3c, the second harmonic is at about 17~dB below the fundamental, corresponding to -17~dBc second-harmonic distortion (SHD). Noting that SHD is our leading term of the total harmonic distortion (THD), our present result falls several dB short of the acceptable THD level for conventional receivers (which is about -20~dBc). However, given the novelty of the atomic receiver and noting that the instrument has not been fully optimized yet, we believe this to be a promising initial result.

\section{Two-tone testing: Intermodulation distortion}\label{Section4TwoToneIMD}

Next we investigate the non-linear response of the atomic receiver for inter-modulation distortion in the presence of an additional interfering signal tone. 
We perform two-tone atomic receiver testing using the setup shown in Figure~\ref{fig:2tone}. In this process, we simultaneously apply $F_{1}$, as done previously, along with a second tone, $F_{2}$, to the atomic receiver, which is configured to operate at a fixed fixed $F_{LO}=9.9376$~GHz. The response of the atomic receiver to the tones and resulting intermodulation products are observed directly at the IF on the SA, where the applied $F_{1}$ and $F_{2}$ tones each get simultaneously demodulated by the Rydberg atoms and reference $F_{LO}$ field to their respective IF at $f_{1}$ and $f_{2}$.  We investigate two different conditions for the applied tones: (1) $\Delta F/F=10^{-6}$, where $\Delta F=F_2-F_1=10$~kHz for $F_1=+100$~kHz, with both tones at approximately the 6~dB IF bandwidth of the receiver (see Figure~\ref{fig:selectivity} and Table~\ref{table:if-bandwidth}) and (2) $\Delta F/F=10^{-4}$, where $\Delta F=F_2-F_1=1$~MHz for $F_1=+100$~kHz, with the second tone at a comparably higher frequency at the 22~dB IF bandwidth, well outside of the resonant line shape shown in Figure~\ref{fig:selectivity}(a).  In all cases the field amplitudes of both tones are equal. 

\subsection{Intermodulation distortion at $\Delta F/F=10^{-6}$}\label{subsec:dff=1e-6}

\begin{figure}[h!]
\centering
\includegraphics[width=0.85\linewidth]{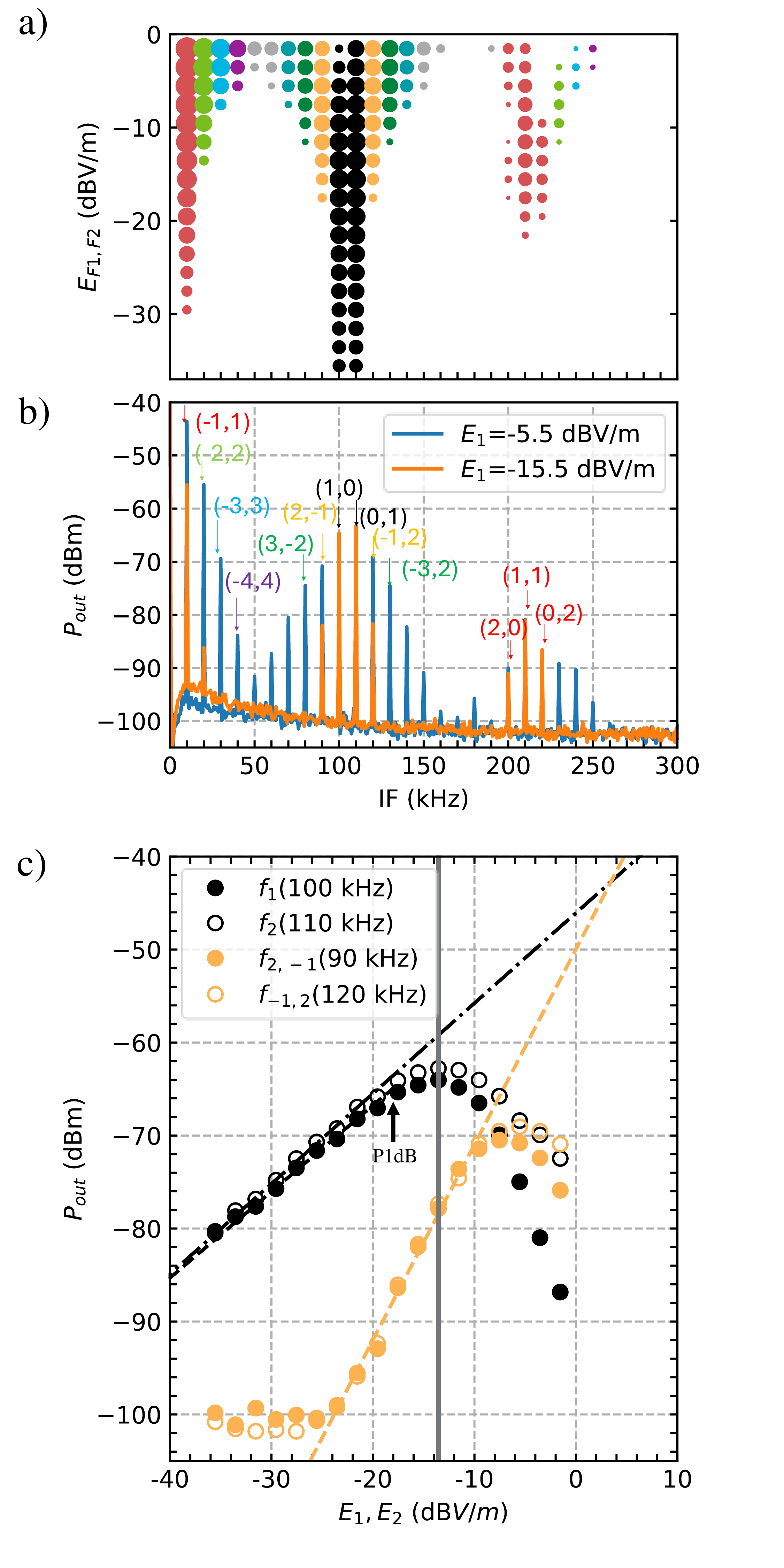}
\caption{Atomic receiver two-tone testing at $\Delta F/F=10^{-6}$: (a)  IMD-map showing the intermodulation peaks in the IF range spanning from 10~kHz to 300~kHz (x-axis) for a selection of applied  RF electric field for the two tones $F_{1}$ and $F_{2}$ (y-axis). The dot size and color scheme are explained in the text. (b) Power spectra of the receiver response in the IF measured on the SA for two tone strengths $F_{1}$=9.93770~GHz and $F_{2}$=9.93771~GHz applied at the vapor cell with electric-field amplitudes $E_{1}=-5.5$~dBV/m (orange) and $E_{2}=-15.5$~dBV/m (blue). This spectra was taken at 1~kHz RBW to accelerate the acquisition with such a large frequency span of 3~MHz, resulting in an elevated noise level. The peak indexing scheme is explained in the text. (c) Receiver IF output power in dBm versus two-tone applied electric field $E_1=E_2$ in dBV/m for the fundamental products (1,0) and (0,1) and third-order IMD (2,-1) at $f_{2,-1}=90$~kHz and (-1,2) at $f_{(-1,2)}=120$~kHz. Linear fits are indicated in dashed lines, yielding the measured P1dB$=-17.5(-15.5)$~dBV/m for $f_1(f_2)$(black arrow), IP3$=2.7$~dBV/m for $f_{-1,2}$ to $f_{1}$, and Figure of Merit FoM=IP3-P1dB=20.2~dB.}
\label{fig:inband}
\end{figure}

To test IMD at $\Delta F/F=10^{-6}$ we set $F_{1}=9.93770$~GHz, where $f_{1}=F_{1}-F_{LO}=100$~kHz and $F_{2}=9.93771$~GHz where $f_{2}=F_{2}-F_{LO}=110$~kHz producing two simulteneous IF responses at  $f_{1}=$100~kHz and $f_{2}=110$~kHz, with a two-tone frequency separation of $\Delta F=F_{2}-F_{1}=10$~kHz and a relative frequency separation of $\Delta F/F=10^{-6}$. Figure ~\ref{fig:inband} shows the results of this two-tone measurement. In (a) we show an IMD-map that is constructed by monitoring each of the detected IMD peaks in the IF range spanning from 10~kHz to 300~kHz (x-axis) for a selection of applied  RF electric field strengths at the atoms' location for the two tones $F_{1}$ and $F_{2}$ (y-axis). Each dot on the IMD-map represents a detected peak on the SA, and the size of the dot indicates the relative signal strength for that peak. The color code of the peaks in the IMD-map is chosen such that the fundamental peaks ($F_{1}$ and $F_{2}$) are black, second-order IMD peaks are red, third-order ones are yellow, fourth-order ones are light-green, fifth-order ones are dark-green, sixth order ones are cyan, seventh order ones are teal, eighth order ones are purple, and peaks that represent IMD of orders higher than 8 are gray. This color notation will be used throughout this paper. Here, we observe that intermodulation starts to develop as the electric field strength increases. The plot in (b) is a cut out of two traces from the map in a), showing the spectral response of the atomic signal for two different electric field strengths of $F_{1}$ and $F_{2}$. The intermodulation peaks are indexed as ($n_{1}$ and $n_{2}$) such the IMD peak order can be identified by the peak frequency in the IF spectra by $f_{n1,n2}=n_{1}\times$$f_{1}+n_{2}\times$$f_{2}$ where the order of the IMD is given by N$=|n_1|+|n_2|$. For example, the first peak labeled (-1,1) represents a second-order IMD ($N=2$) and shows up at $f_{-1,1}=10$~kHz, and the second peak labeled (-2,2) represents a forth-order IMD ($N=4$) and shows up at $f_{-2,2}=20$~kHz. The index scheme will also be used throughout this paper and the IMD peaks will be colored  using the previously described color scheme.  In this plot we can clearly identify the IMD peaks up to the 8$^{th}$ order ($f_{4,-4}$).

The bottom plot in (c) shows the detected signal strength of the fundamental peaks and the two strongest $3^{rd}$ order IMD ($f_{2,-1}=90$~kHz and $f_{-1,2}=120$~kHz) as function of electric field strength. The vertical line in the plot serves as a reference marking the LO field signal strength. The dashed lines represent a linear fit of the data within the range before saturation ($E_1=E_2<-20$~dBV/m). For the fundamental response at $f_{1}$ and $f_{2}$, we observe a near-linear behavior with slopes of $0.97\pm0.03$ and $0.92\pm0.03$, respectively, and a P1dB compression point at -20(-22)~dBm for $f_{1}(f_2)$. The corresponding RF electric fields of the two tones at the P1dB compression points are approximately -17~dBV/m. The 1-dB compression points are determined by solving for where the data point deviates the linear fit by 1~dB.The observed slopes are close to a slope of 1, as expected for fundamental tones in classical receivers. Regarding the selected intermodulation peaks in the plot ($f_{2,-1}=90$~kHz and $f_{-1,2}=120$~kHz), the linear fit indicates a slope of $2.1\pm0.08$, which is suppressed relative to the expected slope of 3 for classical RF systems. We attribute these differences in the atomic heterodyne receiver to the non-linear responses and multi-wave mixing effects present in the quantum Rydberg EIT system, which are anticipated to exhibit deviations from classical RF electronics. From the fit we measured an IP3$=2.7$~dBV/m.
It is imperative to emphasize that these additional peaks are not attributable to any electronics artifact in the detection or on the back-end of the receiver. Rather, they solely arise as the high-order effects of the atoms themselves. In RF heterodyne measurements, the atoms experience the RF electric fields from the LO and the signal (F$_1$ and F$_2$) fields. According to the superposition principle of electromagnetism in free space, the electric-field amplitudes due to multiple field components are a linear sum over all components. As is commonly the case, harmonic generation and IMD arise as a result of nonlinear response of an element (which may be a diode, an amplifier, or a receiver system) at high electric fields. In the present case, it is the nonlinear response of the atomic EIT medium to the electric field that gives rise to harmonic generation and IMD.

%The nonlinear response of the atomic receiver is modeled by time-dependent solution of the Lindblad equation described in the next section. % for the underlying four-level system and suitable spectral analysis of the time-dependent coherence $\rho_{12}(t)$, where subscript 1 refers to the atomic ground state and 2 to the intermediate state of the cascade EIT system. The four atomic levels are coherently coupled by the EIT probe and coupler lasers, the RF LO field, and up to two SIG fields. The model yields output powers at the IF frequencies, their overtones, and the IMD frequencies. A complete description of the model with theoretical results are presented in the supplemental material.

\subsection{Inter-modulation distortion at $\Delta F/F=10^{-4}$}
\begin{figure}[h!]
\centering
\includegraphics[width=0.82\linewidth]{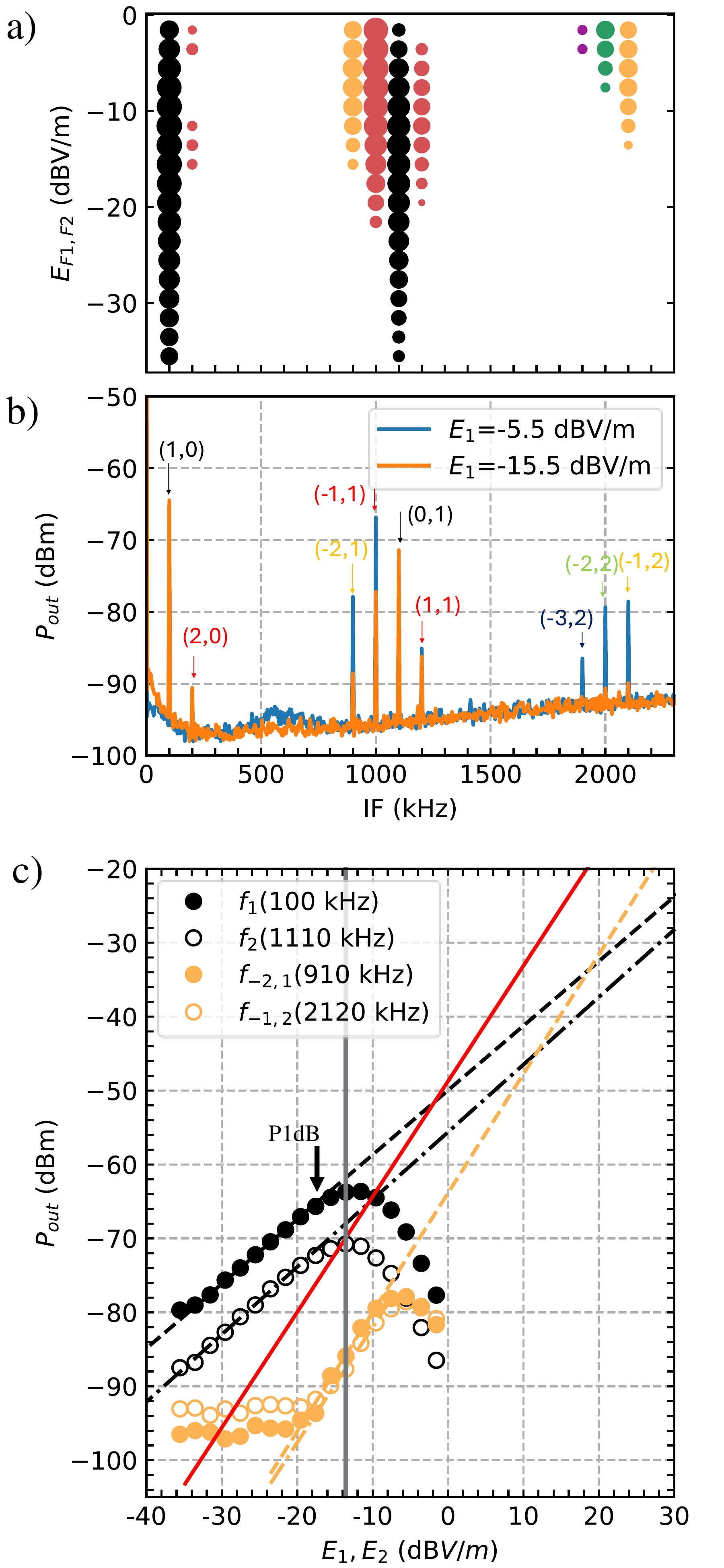}
\caption{
Atomic receiver two-tone testing at $\Delta F/F=10^{-4}$: (a)  IMD-map showing the intermodulation peaks in the IF range spanning from 10~kHz to 2.300~kHz (x-axis) for a selection of applied  RF electric fields for the two tones $F_{1}=9.93770~GHz$ and $F_{2}$=9.93871~GHz (y-axis). (b) Power spectra of the receiver response in the IF measured on the SA for two values of the electric-field amplitude $E_{low}=-15.5$~dBV/m (red) and $E_{high}=-5.5$~dBV/m (orange). The peaks indexing scheme is explained in the text. This spectra was taken at 1~kHz RBW to accelerate the acquisition with such a large frequency span of 3~MHz, resulting in an elevated noise level. (c) Receiver IF output power in dBm versus two-tone applied electric field $E_1=E_2$ in dBV/m for the fundamental products (1,0) and (0,1) and the third-order IMD (-2,1) at $f_{-2,1}=910$~kHz and (-1,2) $f_{-1,2}=2120$~kHz. Here, the RBW is 1~Hz. Linear fits are indicated in dashed lines, yielding the measured P1dB$=-16.5(-15.5)$~dBV/m for $f_1(f_2)$(black arrow), IP3$=19.3(11.9)$~dBV/m for $f_{-2,1}$ to $f_{1}$($f_2$)  , and Figure of Merit FoM=IP3-P1dB $\approx 35$~dB.}
\label{fig:outband}
\end{figure}

We also tested IMD at $\Delta F/F=10^{-4}$ by keeping $F_{LO}=9.93760$~GHz and $F_{1}=9.93770$~GHz, but setting the frequency of the second tone near the end of the receiver's linear response (refer to Figure~\ref{fig:selectivity}), $F_{2}=9.93871$~GHz. This increases $f_{2}$ to 1.110~MHz and leads to an absolute frequency separation of $\Delta F=F_{2}-F_{1}=1.01$~MHz.
Figure~\ref{fig:outband} shows the results for the two-tone measurement with $\Delta F/F=10^{-4}$. Figure~\ref{fig:outband} (a) shows an IMD-map that illustrates all the IMD peaks in the IF band as a function of the $F_1$ and $F_2$ electric field strengths at the atoms' location (as defined previously). In this plot we show that the atomic vapor continues to mix down the applied tones to two intermediate frequencies $f_{1}$ and $f_{2}$. In comparison with the results in Figure~\ref{fig:inband} (a), we observe fewer peaks. 
This difference is a result of the large $\Delta F$, which allows only a few IMD frequencies ($f_{n1,n2}$) within the atomic IF response bandwidth. For instance, for the $\Delta F/F=10^{-6}$ case in Sec.~\ref{subsec:dff=1e-6} the $6^{th}$-order IMD peak $f_{-3,3}$ occurs at 30~kHz, which is well within the atomic IF bandwidth, whereas in the present $\Delta F/F=10^{-4}$ case the same IMD peak is at 3.030~MHz, which is outside of the atomic IF bandwidth (table \ref{table:if-bandwidth}).

In the measurement presented in Figure~\ref{fig:outband}, the two tones were set to have the same electric field strength, but the fundamental tones $f_1$ and $f_2$ differ in signal strength, with the signal at $f_2$ being 10~dB lower than $f_1$. This finding is in good qualitative agreement with Figure~\ref{fig:selectivity}(c), which shows approximately 10~dB drop in IF response between 100~kHz and 1~MHz.  
%is attributed to the finite Rydberg EIT linewidth and reduced atomic response at 1~MHz away from the resonance.
Figure~\ref{fig:outband}(b) shows a selection of two SA traces, displaying the spectral response of the atomic signal for two different electric field strengths of  the signals at $F_{1}$ and $F_{2}$, with the IMD peaks labeled as defined previously. In this plot, we observe IMD products up to the 5$^{th}$ order ($f_{-3,2}=1920$~kHz).

Figure~\ref{fig:outband}(c) shows the detected signal strength of the fundamental peaks and the two strongest $3^{rd}$-order IMDs ($f_{-2,1}=910$~kHz and $f_{-1,2}=2120$~kHz) as a function of electric field strength. The vertical line in the plot marks the field of the LO. The dashed lines show linear fits of the data in the sub-saturation regime $E_1=E_2<-20$~dBV/m; the fits are extrapolated to the right margin of the plot. In the regime $E_1=E_2 \lesssim -20$~dBV/m, we observe a near-linear behavior for the fundamental response at $f_{1}$ and $f_{2}$, with slopes of $0.87\pm0.01$ and $0.91\pm0.02$, respectively, which is near the expected value of 1. The P1dB compression points correspond to RF electric fields of the two tones equal to $-16.5$ and $-15.5$~dBV/m for $f_1$ and $f_2$, respectively. For both $3^{rd}$-order IMD peaks selected in Figure 5~(c), the linear fits have slopes of $1.6\pm0.2$, which is significantly less than our finding in the analogous measurements at $\Delta F/F=10^{-6}$ in Figure~\ref{fig:inband}~(c). We also recall, for comparison, that for $3^{rd}$-order IMD in classical RF systems a slope of 3 would be expected. Again, we attribute this difference to the quantum nature of the Rydberg receiver. 

The IP3 points for the $\Delta F/F=10^{-4}$ 
measurements in Figure~\ref{fig:outband}~(c) are determined to be 19.3~dBV/m and 11.9~dBV/m for $f_1$ and $f_2$, respectively. It is noteworthy that these values are fairly large: they contrast with analogous measurements at $\Delta F/F=10^{-6}$ in Figure~\ref{fig:inband}~(c), where we found an IP3 as low as $2.7$~dBV/m. In view of our finding regarding the slopes of the $3^{rd}$-order IMDs (see previous paragraph), this difference is largely attributed to the comparatively small slope of the 3$^{rd}$-order IMD found for $\Delta F/F=10^{-4}$. A smaller slope of the $3^{rd}$-order IMD pushes the IP3 points farther out. Moreover, for $\Delta F/F=10^{-4}$ [Figure~5~(c)] the IP3 for $f_2$ is about 7~dB lower than for $f_1$. We attribute this behavior to the fact that $f_2$ is near the atomic IF bandwidth, where the atomic response is reduced by about 10~dB relative to the response at $f_1$, pushing the corresponding IP3 to a lower value.

In classical RF systems, IMD is quantified by the power ratio between the fundamental and the $3^{rd}$-order IMD responses. The spur-free dynamic range (SFDR) of a receiver is given by $\frac{2}{3}$(IP3$_{out}$-N$_0$), where N$_0$ is the noise floor of the non-linear or harmonic component~\cite{Pozar.2000} at 1~Hz resolution bandwidth (RBW) on the SA, and IP3$_{out}$ is the output power at the IP3 point. In Figure~\ref{fig:outband}~(c), for $f_1$ and the $3^{rd}$-order IMD we read, on the y-axis, an IP3$_{out}$ of -34~dBm. Also, the SA noise floor was N$_0$=-121~dBm (not shown in the figure). We therefore determine a SFDR of the atomic receiver of $\frac{2}{3}$(IP3-N$_0$)=58~dB.

To benchmark the IMD performance of the atomic quantum receiver to that of a typical classical receiver, we introduce a figure of merit (FoM) for distortion, defined as the difference between the IP3 and P1dB compression point as FoM=IP3-P1dB, where IP3 and P1dB are read on the input (x-axis) in Figure~\ref{fig:outband}~(c). This metric quantifies the additional signal strength that would have to be applied to the input of the receiver (whether classical or quantum) before a third-order non-linear response would become significant relative to the receiver's compression point. From Figure~\ref{fig:outband}~(c), we measure a FoM for the atomic receiver of approximately 35~dB, which compares favorably to many classical low-noise amplifiers (LNAs) at 10~GHz. These LNAs typically exhibit a distortion FoM of 12~dB or less (see, for example, commercial devices from Mini-Circuits part No. ZX60-06183LN+ or Pasternack part No. PE15A1032). An illustrative third order IMD response ($f_{-2,1}$) for the a classical LNA with a FoM=12~dB is overlaid in Figure~\ref{fig:outband}~(c) in the red line.
 
There are two important distinctions to keep in mind in the above comparisons. First, the atomic receiver measures OTA free-space RF signal waves, while LNAs process electronic RF signals. Second, not unrelated to the first, the FoM comparison uses a relative metric and does not provide an absolute RF signal metric measurable as a FoM for distortion by both receiver types. Obtaining an absolute signal RF field metric that can be compared for both receiver types would require specific assumptions about the antenna aperture or similar transducer used to convert a free-space RF signal field to an analogous the RF electric signal input into an LNA. Since the antenna transducer details vary substantially for different antenna types and applications of interest, it is not incorporated in the comparison presented in our paper. Generally, and without regard to the front-end antenna used, the atomic quantum receivers can exhibit a greater tolerance to nearby interference signals compared to classical LNAs within their respective dynamic ranges \cite{reed2014lna}.

\section{Nonlinear response Modeling}

The nonlinear response of the atomic receiver is modeled by time-dependent solution of the Lindblad  master equation for the underlying four-level system. Indices 1, 2, 3 and 4 refer to ground state, excited state, $nD$-Rydberg state and RF-coupled Rydberg state, respectively. We numerically find time-dependent solutions for the ground-excited-state coherence, $\rho_{1,2} (t)$. Spectral analysis of $\rho_{1.2}$ then yields the signals, harmonics and mixing products.

\begin{equation}\label{master_equation}
\dot {\rho } = \frac{{\partial \rho }}{{\partial t}} =  - \frac{\rm{i} }{\hbar }\left[ {{\mathbf{H}},\rho } \right] + \mathcal{L}
\end{equation}
where $\mathbf{H}$ represents the Hamiltonian:
\begin{equation}
H = \hbar 
\begin{pmatrix}
0 & \Omega_{1,2}/2 & 0 & 0 \\
\Omega_{1,2}^*/2 & \delta_{2} & \Omega_{2,3}/2 & 0 \\
0 & \Omega_{2,3}^*/2 & \delta_{3} & \Omega_{3,4}/2 \\
0 & 0 & \Omega_{3,4}^*/2 & \delta_{4} 
\end{pmatrix},
\end{equation}
where $\delta_{n}$ are the dressed-state energies,
\begin{eqnarray}
\delta_1 = & 0 \nonumber \\
\delta_2 = &  - \Delta_P + k_P v  \nonumber \\
\delta_3 = & \, \delta_2 - \Delta_C - k_C v  \nonumber \\
\delta_4 = & \delta_3 - \Delta_{RF}
\end{eqnarray}
where $v$ is the atom velocity along the laser beam directions, $\Delta_P$, $\Delta_C$ are the probe and coupler detunings from the respective atomic transitions at zero velocity. The $k_*$ are corresponding wavenumbers, which are defined positive. $\Delta_{RF}$ is the detuning of a selected RF reference frequency, $\omega_{ref}$, from the Rydberg microwave transition frequency. One may chose $\Delta_{RF}=0$. 

The four atomic levels are coherently coupled by the EIT probe and coupler lasers, the RF LO field, and up to two SIG fields. The $\Omega_{n,m}$ are the complex Rabi frequencies between levels $n$ and $m$. The net RF Rabi frequency depends on the angular frequency differences $\eta_{LO}$, $\eta_1$ and $\eta_2$ relative to the selected RF reference frequency, $\eta_{LO} = \omega_{LO} - \omega_{ref}$ etc., 
\begin{equation}
\Omega_{3,4} = \Omega_{LO} \exp (- {\rm{i}} \eta_{LO} t) 
+ \Omega_{1} \exp (- {\rm{i}} \eta_1 t)
+ \Omega_{2} \exp (- {\rm{i}} \eta_2 t) \quad.
\end{equation}
There, the fixed complex Rabi frequencies $\Omega_{LO}$,
$\Omega_{1}$ and $\Omega_{2}$ for LO, SIG 1 and SIG 2 may incorporate additional fixed phases.
% Maehhh .... and $\rho$ is the density matrix:
%\begin{equation}
%\rho = \begin{pmatrix}
%\rho_{11} & \rho_{12} & \rho_{13} & \rho_{14} \\
%\rho_{21} & \rho_{22} & \rho_{23} & \rho_{24} \\
%\rho_{31} & \rho_{32} & \rho_{33} & \rho_{34} \\
%\rho_{41} & \rho_{42} & \rho_{43} & \rho_{44} 
%\end{pmatrix},
%\end{equation}
%where subscript $n$ and $m$ refers to all the 4 atomic states involved in this excitation (1-ground, 2-intermediate, 3-Rydberg nD state and 4-RF coupled Rydberg state). 

The Lindblad operator $\mathcal{L}$ accounts for all atomic decays and decoherence processes, which are the natural decay of the excited state, state-dependent collisional dephasings for all four levels, and a global atom recycling rate related to the inverse of the typical atom-field interaction time.     

The time-dependent complex absorption coefficient for the probe then is
\begin{equation}
\alpha_{1,2} (t) = \frac{4 \pi d^2 n_V }{\hbar \epsilon_0 \lambda_P \vert \Omega_{1,2} \vert} \int dv P(v) \rho_{12} (v,t)  \quad ,
\end{equation}
with the normalized one-dimensional Maxwell distribution $P(v)$, the $m$-averaged dipole moment $d = 1.99 e a_0$ for the Cs D2 line,  the atom density $n_V$ in units m$^{-3}$, and the probe wavelength $\lambda_P$.  

We then obtain the imaginary parts of the complex Fourier transform of 
$\alpha_{1,2}$, denoted $\tilde{\alpha}_{1,2} (\omega_{IF})$, where the IF frequency domain of $\omega_{IF}$ covers both positives and negatives. 
Note that $\tilde{\alpha}_{1,2}$ is real-valued in this definition.
The receiver signal power at a $ \vert \omega_{IF} \vert $, denoted $P_{out} (\omega_{IF})$, then equals the sum of the squares of positive- and negative-frequency amplitudes, $\tilde{\alpha}_{1,2} (\omega_{IF})$ and $\tilde{\alpha}_{1,2} (-\omega_{IF})$, with a positive $\omega_{IF}$. Note that in this formulation the argument $\omega_{IF}$ can be any frequency in the IF band, including the nominal IF frequencies of the signals, $| \omega_{LO} - \omega_1|$ and $| \omega_{LO} - \omega_2|$, overtones of these, as well as IMD products.

%GEORG: please add a paragraph explaining how the three fields (LO and SIG1 and SIG2) are added into the matrix/model - done above

\begin{figure}[ht!]
\centering
\includegraphics[width=0.85\linewidth]{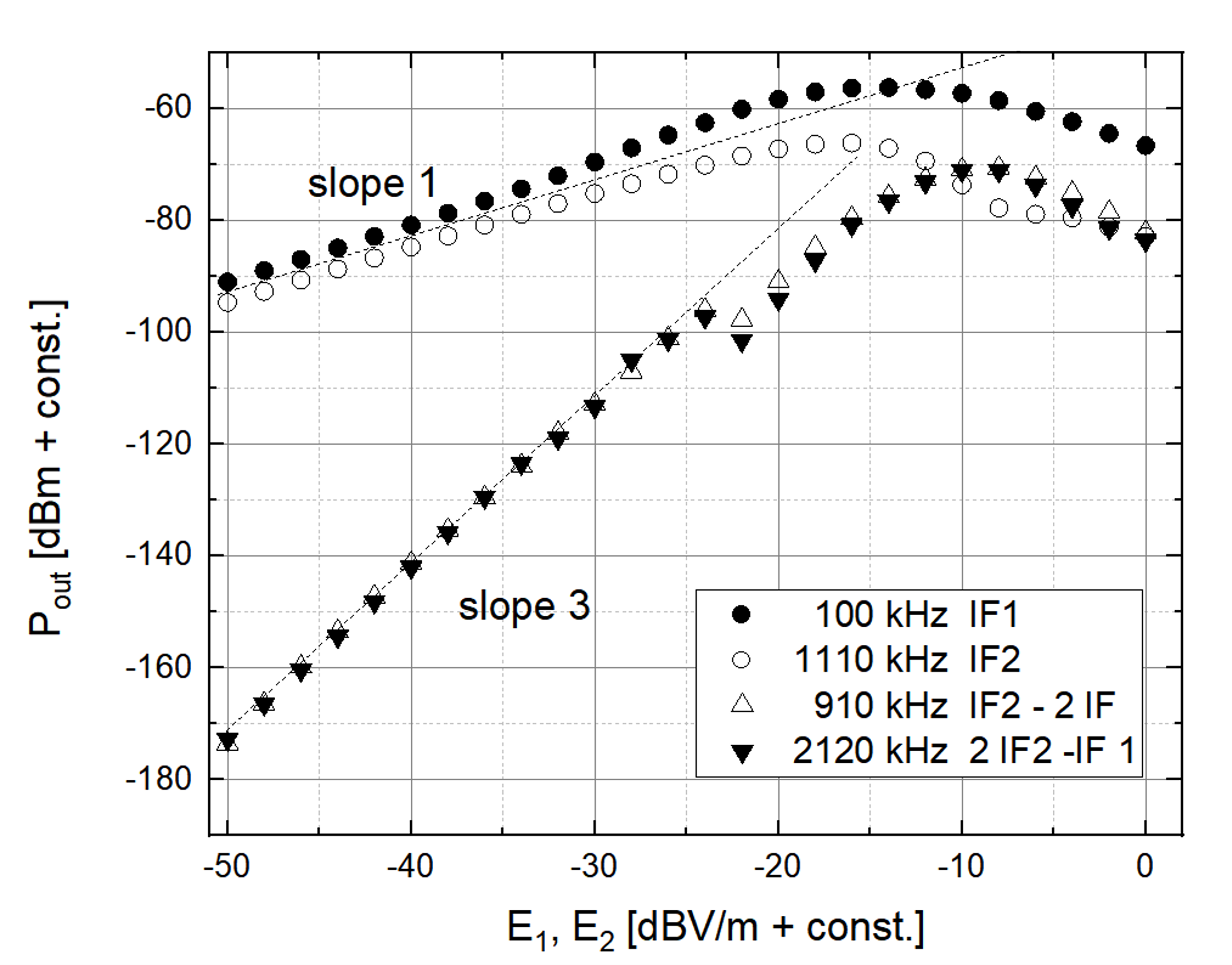}
\caption{Simulation result for parameters similar to those used in the experimental Figure~5. Symbols show simulated data. Lines with the indicated fixed slopes have been added into the plot to estimate the slopes of the data. Circles show experimental data obtained for the IF signals. IF and selected IMD frequencies are indicated in the legend. It is seen that both IF signals have a slope near 1. The lower IF frequency has the stronger response, and the higher IF frequency rolls off slightly before the lower IF. These observations closely match the experimental data in Figure~5 of the main text. The triangles show the third-order IMD products indicated in the legend. The IMD products have a slope near 3 and roll off at about 5~dBV/m above the roll-off fields of the IF signals, which also is in qualitative agreement with Figure~\ref{fig:outband} in the main text. The simulation does not include noise.
}
\label{fig:sims}
\end{figure}

The model yields output powers at the IF frequencies, at their overtones, and at the IMD frequencies. While additional details of the simulation work are beyond the scope of the present paper, in Figure~\ref{fig:sims} we show simulated results on third-order IMD products for a selected case. The simulated input field is given in units of dBV/m and the output power in units of dBm. In the computation, the SIG fields are entered in the form of respective Rabi frequencies, which are proportional to the SIG electric fields. The proportionality constant, which includes the Rydberg transition's electric dipole moment, is accounted for by an additive constant on the dBV/m scale. The receiver output $P_{out} (\omega_{IF})$ is modeled as described above.
%via the modulation of the atomic absorption coefficient at the frequencies of interest, 
The $\omega_{IF}$-values of interest include the fundamental IF frequencies, their harmonics, and various IMD frequencies. Multiplicative constants that arise from vapor-cell, laser-beam and gain parameters are accounted for by an additive constant on the dBm scale.

The model clearly shows the rise of high-order harmonics and intermodulation peaks as the RF signal power is increased. The simulation result shown here in Figure~\ref{fig:sims} is in good qualitative agreement with the corresponding measured data in Figure~\ref{fig:outband} of the main text. Notably, the slopes and roll-off characteristics are a good match for both the IF and the IMD products. The simulation does not include noise on the signal and inhomogeneities of both the optical and RF fields. Nonetheless, the presented simulations produce good initial results towards future modeling work and refinement.

\section{Discussion}\label{Section5Discussion}
The non-linear behavior of atomic receivers described in this work are notably unlike those of classical LNAs and electronic systems in that they incorporate an atom-field interaction that is ultimately tied to the invariable atomic structure and fundamental physical constants. Moreover, the types of EIT schemes\cite{schaffer2023multi,eit2022qc,pfau2012eit}, the intensities of the optical EIT-readout beams\cite{sapiro2020time,su2022optimizing,simon2016cavity,shaffer2017cavity,mao2024cavity}, and the atomic vapor-cell conditions play a role\cite{fan2015cell,wu2024dependence,song2018field}. As such, unlike their classical counterparts, the non-linear behaviors of atomic receivers are reproducible under user-determined conditions and can provide unique, controllable RF signatures on the receive-side for predetermined signal sources.  This can be exploited to realize transmit-receive communications modalities for secure communications, RF fingerprinting, and other applications.  As a basic example implementation, a physical encryption scheme may be employed using an atomic receiver in which the non-linear response of a two-tone local oscillator (e.g. output shown in Figure~\ref{fig:inband} can be switched between discrete spectra of higher-order products (e.g. by switching LO amplitudes and frequencies).  The switching between spectra can serve as a coding algorithm to, for example, provide a physical frequency hop table selector in the receiver for an incident RF waveform with FHSS modulation.  

The slopes of the second and third harmonics are close to 2 and 3, the values expected for semiconductor-based systems. This similarity is noteworthy because the atomic receiver differs greatly from those, yet we find similar scalings for the second and third harmonics. We further use Fig. 3 to extract the second- and third-order intercepts, IP2 and IP3. These values commonly serve as measures to rate the non-linearity of receivers. Higher intercepts are desired because they correspond with lower nonlinearity. The value of IP2 is relevant because the second harmonic often is the leading distortion (as is the case in our Fig. 3). The value of IP3 has an elevated importance because it also is an approximate measure for third-order IMDs (see Section V). The latter can be close in frequency to a (fundamental) IF and therefore hard to filter out. In Fig. 3 we find nearly-coincident IP2 and IP3-values of E1 $\approx$ 12.5~dBV/m. For the atomic receiver in Fig.~3, the IP3 value is about 25~dB above P1dB, which is somewhat higher than the typical difference of 10~dB to~15 dB in conventional amplifiers. The 10~dB advantage observed in the atomic receiver likely reflects the fundamental differences in the principles of operation of atomic and conventional receivers. We also note that in conventional systems the difference IP2-P1dB ranges between about 15~dB to 40~dB.  In our atomic receiver IP2-P1dB is about 25~dB, which is in line with conventional systems.

\section*{Conclusion}\label{Section6Conclusion}
In our study, we have presented a first comprehensive characterization of harmonic and inter-modulation distortion behavior in atomic receivers. Through two-tone testing conducted in the SHF band, utilizing tones resonant to an Autler-Townes transition between Rydberg atoms in the atomic-receiver medium, we have observed IMD of orders higher than 6, measured the dependence of the harminics and IMD peaks with incident RF power, obtained crucial metrics such as P1dB and IP3 points, and characterized the dynamic range and spur-free dynamic range of the system. Our results have unveiled a remarkable suppression of harmonic and IMD in atomic receivers under specific operating conditions, diverging from the behavior one would expect for classical receivers. Our observations highlight the unique and advantageous properties of atomic receivers in mitigating nonlinear effects. 
While our experimental results are derived from a specific atomic receiver implementation, many of the observed phenomena, such as the presence and scaling of harmonic and intermodulation distortions, the behavior of the intermediate frequency bandwidth, and the suppression of nonlinearities, are fundamentally rooted in atomic and quantum mechanical processes. These processes are largely universal and therefore expected to generalize across similar atomic receiver systems. Although quantitative parameters like P1dB, IP3, and SFDR may vary with experimental configurations (e.g., laser parameters, cell geometry, and local oscillator power), the underlying mechanisms and trends we describe should be broadly applicable. As such, the insights offered in this work serve not only to characterize a specific device but also to inform the design and interpretation of behavior in a wider class of atomic receivers. %Research on atomic receivers with a focus on signal-processing methods, including works presented in \cite{gong2025rydbergatomicquantumreceivers, gong2025rydbergatomicquantumreceivers2,gong2025rydbergatomicquantumreceivers3}, may in the future be adapted for continued studies of non-linearities in atomic receivers.
In addition, we have described the nature of non-linear behaviors exhibited by atomic receivers and have proposed their exploitation in various applications, including secure signal reception and innovative communication schemes. The findings in our work pave the way for the development of robust and versatile atomic receiver technologies with wide-ranging applications.

\bibliography{References}% Produces the bibliography via BibTeX.

\end{document}